\documentclass[pra,floatfix,shwopacs,amsmath,amssymb]{revtex4}
\usepackage{graphics}

\begin{document}

\newcommand{\one}{\ensuremath{\hbox{$\mathrm I$\kern-.6em$\mathrm 1$}}}
\newcommand{\ad}{\ensuremath{a^\dagger}}
\newcommand{\bd}{\ensuremath{b^\dagger}}
\newcommand{\cd}{\ensuremath{c^\dagger}}
\newcommand{\dd}{\ensuremath{d^\dagger}}
\newcommand{\ed}{\ensuremath{e^\dagger}}
\newcommand{\ket}[1]{\ensuremath{\left|{#1}\right\rangle}}
\newcommand{\bra}[1]{\ensuremath{\left\langle{#1}\right|}}
\newcommand{\braket}[1]{\ensuremath{\left\langle{#1}\right\rangle}}

\title{Spin dynamics for bosons in an optical lattice}

\author{Juan Jos\'e Garc\'{\i}a-Ripoll}
\address{Max-Planck-Institute for Quantum Optics, Hans-Kopfermann-Str. 1,
  Garching, D-85748, Germany}
\author{Juan Ignacio Cirac}
\address{Max-Planck-Institute for Quantum Optics, Hans-Kopfermann-Str. 1,
  Garching, D-85748, Germany}

\date{\today}

\begin{abstract}
  We study the internal dynamics of bosonic atoms in an optical lattice. Within
  the regime in which the atomic crystal is a Mott insulator with one atom per
  well, the atoms behave as localized spins which interact according to some
  spin Hamiltonian. The type of Hamiltonian (Heisenberg, Ising), and the sign
  of interactions may be tuned by changing the properties of the optical
  lattice, or applying external magnetic fields. When, on the other hand, the
  number of atoms per lattice site is unknown, we can still use the bosons to
  perform general quantum computation.
\end{abstract}

\maketitle

\section{Introduction}

In the last years there has been a great development of the storage and
manipulation of cold atoms in optical lattices \cite{Bloch1,Bloch2}. The
interest of these atomic systems is multiple. First, the optical lattices are
accurately described by the Bose--Hubbard Hamiltonian, and exhibit a quantum
phase transition from a superfluid to a Mott--insulator \cite{Jaksch1}.  The
jump from superfluid to insulator may be explored in the experiments in a
time-dependent way \cite{Bloch1}, opening a wonderful playground for the study
of time dependent phase transitions, establishment of coherence, and many other
complex phenomena. Second, as it was shown in Refs. \cite{Duan},
the bosons in the optical lattice may be used to simulate different spin
Hamiltonians, with the advantage over magnetic materials that the parameters
may be changed faster and more easily. Finally, these systems are also a good
candidate for performing scalable quantum computations \cite{Jaksch2,Zanardi}.

In this paper we focus on two applications of optical lattices: quantum
simulation of spin $s=\frac{1}{2}$ Hamiltonians and universal quantum
computing. Our proposals require tools which already belong to current
experiments: (i) an optical lattice with one or more atoms per site, (ii) atoms
with two accessible hyperfine levels, (iii) Raman lasers that connect these
levels, and (iv) a magnetic field or an electric field with the appropriate
spatial dependence.  With these tools, and assuming that the lattice is a
Mott-insulator with one atom per lattice site, we generalize the techniques
developed in Ref. \cite{Duan} to consider also these external elements (Rabi
oscillations and magnetic and electric fields), to engineer a large class of
spin Hamiltonians,
\begin{equation}
  \label{H-spin}
  H = \sum_{\braket{i,j}} \left[\lambda^{z}_{i,j}\sigma^z_i\sigma^z_j+
    \lambda^{\perp}_{i,j}
    \left(\sigma^x_i\sigma^x_j+\sigma^y_i\sigma^y_j\right)\right] +
    \sum_i \vec{h}_i\vec{\sigma}_i.
\end{equation}
The flexibility of the system is such, that one may simulate an Ising or
Heisenberg Hamiltonian, on different geometries dictated by the underlying
optical lattice, and even with some randomness in the coefficients
$\lambda_{z,\perp}$ or in the effective magnetic field, $\vec{h}$.

The second focus of this work is quantum computation. The effective spin
Hamiltonians mentioned above could be used to develop a universal quantum
computer, either directly, or by constructing the so called cluster states
\cite{Briegel}.  However, since it is not easy to produce a bosonic crystal
with integer filling factor ---or at least not with the fidelity required for
scalability---, we have developed other methods for quantum computation with
any number of atoms per lattice site. The proposal that we developed in
\cite{prl} and summarize here combines ideas from quantum engineering (for the
definition of qubit and to obtain an effective interaction between them), the
technique of adiabatic passage \cite{holonomias2} (to perform controlled gates
even when the parameters of our system can not be accurately determined), and
ideas from quantum control \cite{Bang} (to cancel dynamical phases in the
adiabatic passage).

The paper is structured in two parts. In the first part we concentrate on the
dynamics of optical lattices in the Mott insulator regime. The lattice will be
loaded with one atom per site, and each atom has two accessible internal state,
so the atoms may be formally identified with $s=\tfrac{1}{2}$ spins. We will
write down the most general Hamiltonian that takes into account the hopping and
interaction of atoms, as well as the influence of external magnetic and
electric fields on the atoms. By adjusting the optical lattice and the external
fields, we will be able to develop an effective Hamiltonian in the spins
representation which covers all possibilities shown in Eq.  (\ref{H-spin}). We
determine the accuracy of this description analytically and numerically, and
briefly study the application of these techniques to generate entanglement
between the atoms. In the second part of the paper we will demonstrate that, if
there are imperfections in the loading of the lattice, the Hamiltonian of the
system is not known with enough accuracy to do quantum computing in a
``traditional'' way. Next we develop a technique to circumvent our ignorance
about the Hamiltonian, using adiabatic passage with the different parameters of
our Hamiltonian to produce a universal set of gates. Finally we estimate the
errors of our proposal, studying both the influence of the speed of the
adiabatic process, and of imperfections in the setup.

% ------------------------------------------------------------

\section{Simulation of spin Hamiltonians}

% ------------------------------ El modelo ------------------------------

\subsection{The Bose-Hubbard model}
\label{sec-bh}
In an optical lattice, pairs of laser beams create a stationary wave which the
atoms see as a periodic potential. If the energies involved in the dynamics are
so small that the second Bloch band never gets populated, we may use the
Bose-Hubbard Hamiltonian to describe the atomic ensemble \cite{Jaksch1}. We
will assume that the lattice is populated with a single atomic species, and
that only two of the hyperfine levels of these atoms may be excited. Then the
Hamiltonian reads
\begin{subequations}
  \label{H-orig}
\begin{eqnarray}
  H &=& \sum_{\langle i,j \rangle} H_{hop}^{(i,j)} + \sum_i
  \left[ H_{int}^{(i)} + H_{mag}^{(i)} + H_{el}^{(i)} + H_{las}^{(i)}\right]\\
  H_{hop}^{(i,j)} &=& -J_a(\ad_{i}a_j + h.c.) + J_b(\bd_{i}b_j +
  h.c.),\\
  H_{int}^{(i)} &=& \tfrac{1}{2}U_{aa}\ad_i\ad_ia_ia_i +
  \tfrac{1}{2}U_{bb}\bd_i\bd_ib_ib_i + U_{ab}\ad_i\bd_ib_ia_i\\
  H_{mag}^{(i)} &=& \epsilon_k (\ad_i a_i - \bd_i b_i),\label{H-mag}\\
  H_{el}^{(i)} &=& \gamma_k (\ad_i a_i + \bd_i b_i),\label{H-el}\\
  H_{las}^{(i)} &=& \frac{\Omega}{2}(\ad_i b_i e^{i\phi} + \bd_i a_i e^{-i\phi}).\label{H-las}
\end{eqnarray}
\end{subequations}
The operators $a$ and $b$ are bosonic destruction operators for atoms in two
degenerate hyperfine states; the indices $i$ and $j$ run over the lattice
sites, and $\braket{i,j}$ denotes a pair of neighboring sites. The constants
$J$ and $U$ depend on the depth of the optical lattice, and they measure the
amplitude of probability of atoms hopping to neighboring sites ($H_{hop}$) and
their effective on-site interaction ($H_{int}$), respectively. The Hamiltonians
$H_{mag}$ and $H_{el}$ account for all possible energy shifts on the internal
states of the atoms.  They may be generated by means of magnetic fields ($a$
and $b$ represent two different hyperfine states which suffer different Zeeman
shifts), or with highly detuned laser beams which induce a Stark shift.
Finally, $H_{las}$ models transitions between the two internal states of the
atoms. The Rabi frequency $\Omega$ and the phase $\phi$ are related to the
intensity and the phase of the laser which induces these transitions.

Throughout the paper we will assume that the system is in the Mott-insulator
regime, $U\gg J$, in which the hopping represents a small perturbation with
respect to all other term of the Hamiltonian.  For the external fields,
$\epsilon_k$ and $\gamma_k$, we will require that they have a simple spatial
dependence, i.~e. they may increase or decrease linearly along a given spatial
direction, as in $\epsilon_k = \epsilon_0 + \delta \times k$. No particular
boundary conditions are imposed, and all results may be trivially generalized
to any geometry of the optical lattice.

It is worth remarking here that, nowadays in most experiments there exists a
residual harmonic potential which is used to further confine the atoms within
the optical lattice. If the gradient of this potential is extremely small
compared to $J$ and $U$, it may be regarded as a spatially dependent
contribution to $\epsilon_k$. In general, however, the harmonic confinement
influences greatly the ground state of the Hamiltonian (\ref{H-orig}) creating
coexisting regions of superfluid and insulator phases \cite{Jaksch1} which are
useless for our purpose. It would be possible to get rid of this potential if
we use additional optical elements to create a barrier that prevents the atoms
from escaping through the borders of a lattice.

\subsection{Effective spin Hamiltonians}
\label{sec-effective}
It is well known that for a Mott insulator made of one bosonic species the
Bose-Hubbard model is equivalent to the XY model, where spin states are
identified with holes and particles in the lattice \cite{Bruder93}. More
precisely, if the filling factor \footnote{Mean number of particles per site.}
is in between $n$ and $n+1$ and the on-site interaction is strong, then the
occupation number of each site will be either $\ket{n}$ or $\ket{n+1}$. We
identify these states with spin states $\ket{+\tfrac{1}{2}}$ and
$\ket{-\tfrac{1}{2}}$, to obtain the effective Hamiltonian
\begin{equation}
  H_{e} =  -\lambda_\perp
  \sum_{\braket{i,j}} (\sigma^x_i\sigma^x_j +
  \sigma^y_i\sigma^y_j) + \sum_k (\epsilon_k+\gamma_k) \sigma^z_i.
\end{equation}
The magnetic interaction, $\lambda_\perp = J (\bar{n}+1)$, is provided directly
by the hopping terms, $H_{hop}$, and it can be rather intense. However, from
the point of view of quantum simulation this approach has an important
restriction: since the number of particles is conserved, the total spin of the
system is constant, and we cannot introduce terms which are proportional to
$\sigma^x$ or $\sigma^y$ in the Hamiltonian.

We can simulate a larger family of spin Hamiltonians if we populate the
lattice with a single atomic species with two degenerate internal states. We
will assume that the lattice is loaded with exactly one atom per site, and
identify the two possible states of each site, $\ad_k\ket{vac}$ and
$\bd_k\ket{vac}$, with the two polarizations of the spin,
$\ket{\pm\tfrac{1}{2}}_k$. The states with single occupation are separated by
an energy gap of order ${\cal O}(U)$ from any other configuration of the
lattice. Since we are deep in the Mott insulator regime, $J \ll U$, states
with double or higher occupation are highly improbable, and we may treat the
hopping term, $H_1=H_{hop}$, as a perturbation with respect to other
contributions, $H_0 = H_{int} + H_{mag} + H_{el} + H_{las}$. Using second
order perturbation theory \cite{Cohen} we will write an effective Hamiltonian
within the spin space, which looks as follows
\begin{equation}
  \label{H-eff}
  \braket{i|H_{e}|j} = \braket{i|H_0+H_1|j}
  +\frac{1}{2}\sum_\xi \braket{i|H_1|\xi}\left[
    \frac{1}{E_i-E_\xi}+\frac{1}{E_j-E_\xi}\right]
  \braket{\xi|H_1|j}.
\end{equation}
While the indices $i,j$ run over the Hilbert space of the spins, $\xi$
represents any configuration with an excess or defect of particles in any well.
The numbers $E_i$, $E_j$ and $E_\xi$ are the unperturbed energies of these
states: $E_i= \braket{i|H_0|i}$, etc.

We will analyze separately how the different terms in Eq. (\ref{H-orig})
influence the effective Hamiltonian (\ref{H-eff}). First of all, if there are
no external fields ($\epsilon_k=\gamma_k=\Omega=0$), $H_e$ is a Heisenberg
Hamiltonian (\ref{H-spin}) with constants given by \cite{Duan}
\begin{equation}
  \lambda_z =
  \frac{J_a^2+J_b^2}{2U_{ab}}-\frac{J_a^2}{U_{aa}}-\frac{J_b^2}{U_{bb}},\quad
  \lambda_\perp = -\frac{2J_aJ_b}{U_{ab}},\quad
  h_z = \frac{J_b^2}{U_{bb}}-\frac{J_a^2}{U_{aa}},\quad h_x=h_y=0.
\end{equation}

If we switch on a linearly growing electric field $H_{el}$ (\ref{H-el}) with
$\gamma_k = \gamma_0 + \delta \times k$, the effective Hamiltonian remains the
same, but the constants change:
\begin{equation}
  \label{lambda-el}
  \lambda_z =
  \frac{(J_a^2+J_b^2)U_{ab}}{2(U_{ab}^2-\delta^2)}-
  \frac{J_a^2U_{aa}}{(U_{aa}^2-\delta^2)}-\frac{J_b^2
  U_{bb}}{(U_{bb}^2-\delta^2)},\quad
  \lambda_\perp = -\frac{2J_aJ_b U_{ab}}{U_{ab}^2-\delta^2},\quad
  h_z = \frac{J_b^2U_{bb}}{U_{bb}^2-\delta^2}-\frac{J_a^2U_{aa}}{U_{aa}^2-\delta^2}.
\end{equation}
Effectively, the application of the electric field $H_{el}$ is equivalent to a
change of the interaction constants, of the form
\begin{equation}
  U_{uv} \to \frac{U_{uv}^2-\delta^2}{U_{uv}},\quad u,v\in\{a,b\}.
\end{equation}
This effect may be used to intensify the interactions, making the spins evolve
faster. But it also may be used to introduce some randomness in the system. For
instance, if $\gamma_k$ does not grow linearly, but fluctuates from site to
site, the constants, $\{\lambda_z,\lambda_\perp,h\}$ will also
fluctuate from site to site, with an expression given by Eq. (\ref{lambda-el})
with $\delta \to \gamma_{i+1}-\gamma_{i}$. 

Another interesting effect is provided by a linearly growing magnetic field
(\ref{H-mag}), $\epsilon_k = \epsilon_0 + \delta \times k$. This contribution
to the Hamiltonian breaks the degeneracy between the states
$\ket{+\tfrac{1}{2},-\tfrac{1}{2}}$ and $\ket{-\tfrac{1}{2},+\tfrac{1}{2}}$.
If the gradient of the magnetic field is weak compared to the interaction
($J\sim\delta \ll U$), the system is still described by a Heisenberg
interaction (\ref{H-spin}), with constants given by Eq. (\ref{lambda-el}),
except for the magnetic field
\begin{equation}
  \label{h-mag}
  h_{z,i} = \epsilon_i +
  \frac{J_b^2U_{bb}}{U_{bb}^2-\delta^2}-\frac{J_a^2U_{aa}}{U_{aa}^2-\delta^2}.
\end{equation}
If the gradient of the magnetic field is comparable to the interaction ($\delta
\simeq U/20$, for instance), the splitting between
$\ket{+\tfrac{1}{2},-\tfrac{1}{2}}$ and $\ket{-\tfrac{1}{2},+\tfrac{1}{2}}$
becomes so large, that they may no longer be connected by the Hamiltonian.
Applying the rotating wave approximation one finds an Ising Hamiltonian,
\begin{equation}
  \label{H-eff-2}
  H_{e} = \sum_i \left[\lambda_z \sigma^z_i\sigma^z_{i+1}+
    h_{z,i} \sigma^z_i\right],
\end{equation}
where $\lambda_z$ is still given by Eq. (\ref{lambda-el}), and $h_i$ by Eq.
(\ref{h-mag}). Another way to achieve an Ising Hamiltonian which was shown in
Ref. \cite{Duan} consists in tuning the properties of the optical lattice so
that $J_b \to 0$. This way we set $\lambda_\perp = 0$ in Eq.  (\ref{H-spin}),
but there is a residual magnetic field, $h_z=-J_a^2/U_{aa}$.

All Hamiltonians which we have shown up to now commute with the operator $S_z
= \sum\sigma^z_i \propto N_a - N_b$. In order to change this component of the
spin, we have to allow transitions between the internal states $a$ and $b$.
Such processes are modeled by the Hamiltonian $H_{las}$ (\ref{H-las}), and
with little work it can be shown that these terms translate into an effective
magnetic field along the X and Y directions of the spin, $\sum_i
(h_x\sigma^x_i + h_y\sigma^y_i)$, with
\begin{equation}
  h_x = \frac{\Omega}{2}\cos(\phi),\quad h_y = \frac{\Omega}{2}\sin(\phi).
\end{equation}

This last term completes all what is required to simulate the family of
Hamiltonians presented in the introduction (\ref{H-spin}). We may now
particularize the previous results for current experiments with Rubidium. If
we do not use state-dependent optical lattices (i.~e. all atoms see the same
potential), we can approximately take $U_{aa}=U_{bb}=U_{ab}=U$, and
$J_a=J_b=J$.  Then, the effective Hamiltonian with electric field Eq.
(\ref{H-spin}) becomes
\begin{equation}
  \label{Heisenberg}
  H_{e} = \sum_i \frac{-J U}{U^2 - (\gamma_i-\gamma_{i+1})^2}
  \vec{\sigma}_i\vec{\sigma}_{i+1},
\end{equation}
and the Ising Hamiltonian with the linearly growing magnetic field is written
as
\begin{equation}
  \label{Ising}
  H_{e} = \sum_i \left[\epsilon_i \sigma_z^i
  - \frac{J^2 U}{U^2-\delta^2}
  \sigma_z^i\sigma_z^j\right].
\end{equation}
In this case, for simplicity, we have omitted corrections which are
proportional to $J^2\delta/(U^2-\delta^2)$, but which only take place at
borders of the lattice.

Of all Hamiltonians which we have shown, the Ising interaction has the
greatest interest for quantum computation, because of its simple form and
because it produces a universal gate known as the controlled-phase gate
\cite{Nielsen}.  However, for many purposes it would be interesting to get rid
of the effective magnetic field which appears both in Eqs. (\ref{H-eff-2}) and
(\ref{Ising}), and which introduces an uncontrolled dephasing. We know two
ways to produce an Ising Hamiltonian without magnetic field. The first one
would be to use Feschbach resonances to increase the interaction between atoms
of type $a$ and $b$, while keeping equal tunneling rates. In other words
$U_{aa}, U_{bb} \ll U_{ab}$ and $J_a=J_b$. If this is the case, then both
$\lambda_\perp$ and $h$ become zero in Eq.  (\ref{H-spin}) and we get the
desired model.  However, Feschbach resonances require intense magnetic fields
which may be difficult to stabilize for a time scale of order
$U_{aa,bb}/J_{a,b}^2$. The other method to get rid of the effective magnetic
field is to set up the conditions which lead to Eq. (\ref{Ising}) and then
apply a spin-echo $\pi$-pulse at times $T/2$ and $T$, where $T$ is the total
duration of the experiment. This way the $k$-th lattice site acquires a phase
$\exp(-i h_k\sigma^z_kT/2)$ during the first half of the experiment, which is
canceled with the unitary operation
$\sigma^x_k\exp(-ih_k\sigma^z_kT/2)\sigma^x_k$ of the second half.

% ------------------------------------------------------------

\subsection{Error bounds}
In deducing the effective models (\ref{H-spin}), (\ref{Heisenberg}), and
(\ref{Ising}) we have performed several approximations. For the Ising model,
the first source of error relates the contributions of
$\sigma_x^{k}\sigma_x^{k+1}$ and $\sigma_y^{k}\sigma_y^{k+1}$. These operators
induce a swap between qubits, so that the state
$\ket{+\tfrac{1}{2},-\tfrac{1}{2}}$ is connected to the
$\ket{-\tfrac{1}{2},+\tfrac{1}{2}}$ and viceversa. Using the interaction
picture and the Born approximation, we can estimate the probability of a swap
happening in a pair of neighboring wells as
\begin{equation}
  \label{swap}
  P^{k}_{\ket{+-}\rightarrow\ket{-+}} =
  \left|\int_0^t d\tau e^{i2\delta}\braket{+-|V|-+}\right|^2
  \leq \frac{4J^4U^2}{\delta^2(U^2-\delta^2)^2}.
\end{equation}

Another possible source of errors which affects all models is the accumulation
of particles on one well. Put in other words, the Hamiltonian (\ref{H-orig})
allows processes in which one particle jumps from one well to a neighboring one
and remains there. To first order in perturbation theory, an upper bound for
this probability is
\begin{equation}
  P^{k}_{i\rightarrow\xi} =
  \left|\int_0^t d\tau e^{i (U-\delta)}\braket{i|V|\xi}\right|^2
  \leq \frac{2J^2}{(U-\delta)^2}.
\end{equation}
This probability is larger than the one for qubit swaping (\ref{swap}), which
means that in the worst case, the error when considering $M$ wells is at most
of order $E = M {\cal O}\left(J^2/(U-\delta)^2\right)$.

The last and most dangerous source of error is the possibility of having in
the initial state a cell with more than one atom. If we are simulating the
Heisenberg model, these extra atoms may hop to neighboring sites. The
scattering of these atoms with those of opposite polarization leads to very
fast changes on the lattice, and our picture of localized spins breaks down.
Therefore it is extremely important in these experiments to begin with a well
prepared Mott phase with filling factor $n=1$.

\begin{figure}
  \centering
  \resizebox{7cm}{!}{\includegraphics{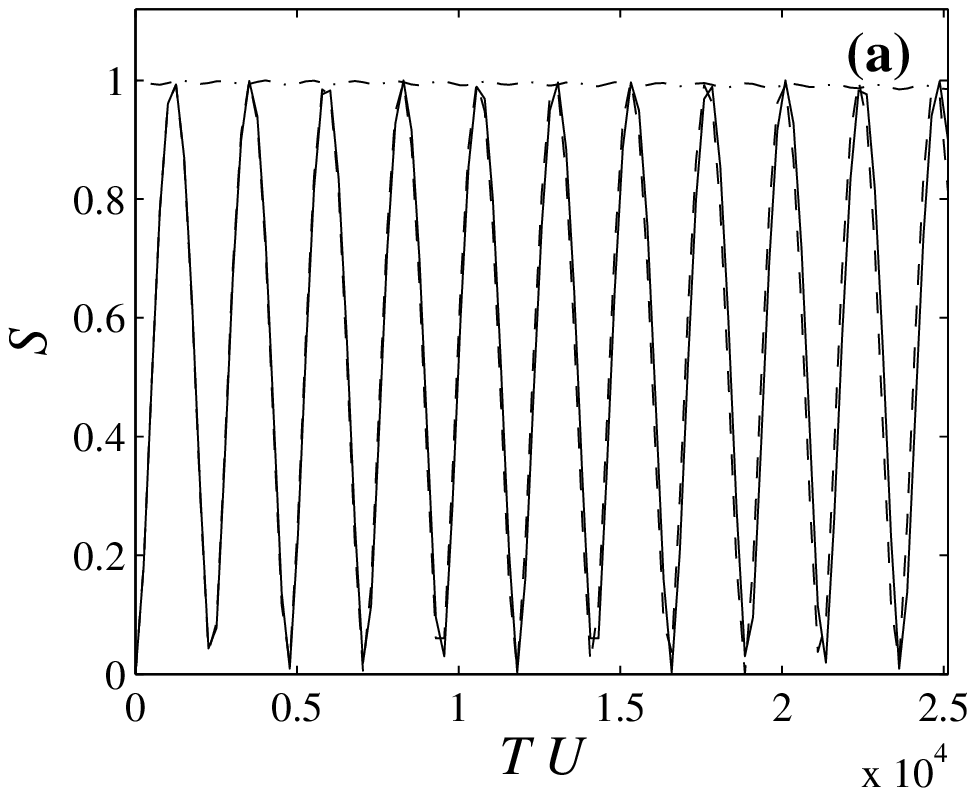}}
  \resizebox{7cm}{!}{\includegraphics{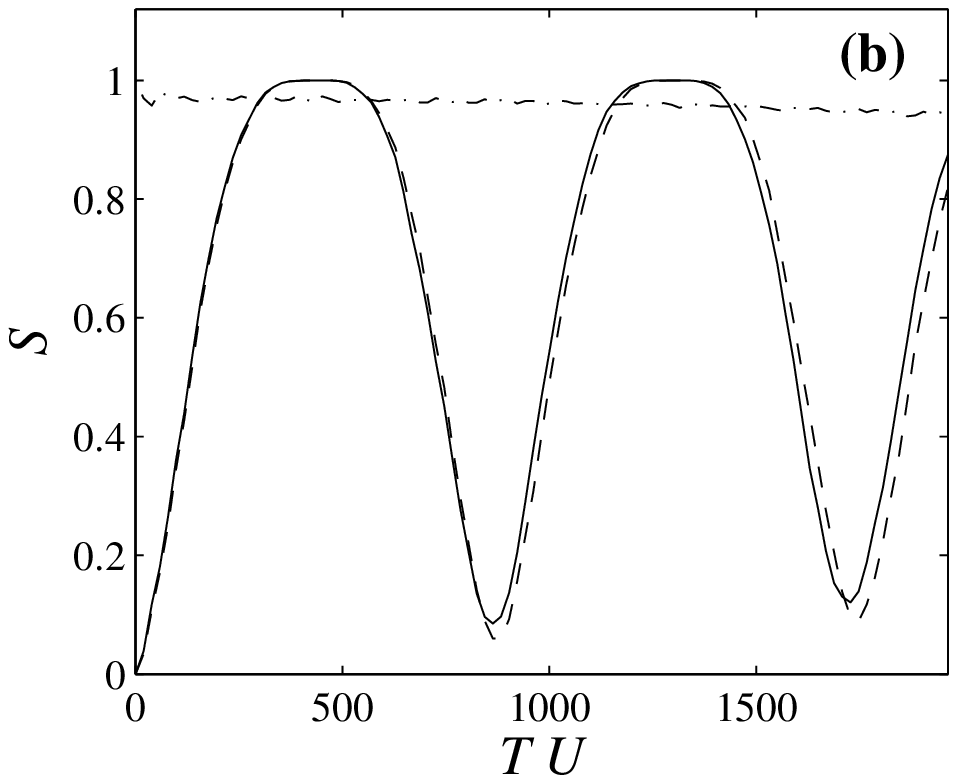}}
  \caption{\label{fig-two}
    Evolution of entanglement as a function of time for an initial state given
    by Eq. (\ref{symmetric-state}), evolving under the conditions which lead to
    the effective Hamiltonian of Eq. (\ref{H-eff-2}). (a) We plot the entropy
    of the reduced density matrix of one well (solid line) out of two, when
    jumping from $J=0$, $\delta\equiv\epsilon_{k+1}-\epsilon_k=0$ to
    $J=0.01U$, $\delta=0.05U$. In dash-dot line we also show
    the fidelity of the operation with respect to the effective Ising
    Hamiltonian (\ref{Ising}).  (b) We perform a similar plot for a system with
    four (dashed line) and five (solid line) lattice sites, for $J=0.04U$ and
    $\delta=0.3U$. In dash-dot line we plot the fidelity of the effective
    Hamiltonian (\ref{Ising}) when describing this evolution.}
\end{figure}

% ------------------------------------------------------------

\subsection{Fidelity}
We have applied our model Hamiltonian (\ref{Ising}) to study how entanglement
is created between two, three and more atoms in an optical lattice. The idea
is not only to perform numerical experiments which show us how much
entanglement can be produced in a realistic experiment, but also to check the
accuracy of our effective Hamiltonian (\ref{H-spin}) when describing the
process.

Our numerical experiments typically proceed as follows. First we prepare an
initial state which has one atom per lattice site, and which we can identify as
a spin state. Then we choose the parameters of the optical lattice
(\ref{H-orig}) so that the theory developed in Sec. \ref{sec-effective}
applies, and simulate the evolution of the whole system considering also the
states with double and higher occupation numbers. Finally we measure the
fidelity of the effective Hamiltonian, $H_e$, as
\begin{equation}
  {\cal F} = |\braket{\psi(0)|e^{i H t} e^{-i H_{e} t}|\psi(0)}|^2,
\end{equation}
where $H$ is the full model depicted in Eq. (\ref{H-orig}).  The deviation of
this quantity from 1 measures the error committed because of trying to
describe the evolution with our simplified model (\ref{H-spin}).

We have performed two sets of simulations in which we check the fidelity of
the Ising and of the Heisenberg Hamiltonians, respectively. In the first row
of numerical experiments the starting point is a product of each atom being
either in state $a$ or $b$
\begin{equation}
  \label{symmetric-state}
  \ket{\psi(0)}= \left[\prod_{k=1}^M \tfrac{1}{\sqrt{2}}
    \left(\ad_k + \bd_k\right)\right]\ket{vac}
  = \bigotimes_{k=1}^{M} \left[\tfrac{1}{\sqrt{2}}
    \left(\ket{+\tfrac{1}{2}}+\ket{-\tfrac{1}{2}}\right)\right]
  = \ket{+\tfrac{1}{2},\ldots,+\tfrac{1}{2}}_x.
\end{equation}
From a physical point of view, this configuration may be prepared by loading
the lattice with one atom per site, switching the tunneling off, $J\simeq 0$,
pumping all atoms into internal state $a$, and finally applying uniformly a
$\pi/2$ laser pulse over the whole lattice. Beginning with the previous state
we customize the optical lattice so that the effective interaction corresponds
to the Ising model (\ref{Ising}).  For the results in Fig. \ref{fig-two}(a) we
have lowered the optical lattice until $J=0.01U$ and set up a magnetic field
with $\delta = 0.05U$. As shown in Ref.  \cite{Briegel}, the ensemble of atoms
should evolve periodically from a product state to an entangled state with
maximal connectivity and large entanglement persistence, also known as
\emph{clusters}. To measure the entanglement of one qubit with respect to the
rest, we compute the von Neumann entropy $S(\rho_k) = \sum
-\lambda_k\log_2(\lambda_k)$, where $\rho_k=\sum_k
\lambda_k\ket{\psi_k}\bra{\psi_k}$ is the reduced density matrix of one well.
In Figure \ref{fig-two}(a) we show how this value evolves for a system of two
wells and two atoms. In order to make the process faster, we have made
numerical experiments with larger hopping strengths, $J=0.04U$, and stronger
field gradients, $\delta=0.3U$, for lattices with four and five wells.  The
results are plotted in Figure \ref{fig-two}(b). The maximum entropy is reached
in all wells simultaneously for state which is close to a cluster. But now we
notice that the process is not perfect. First, the due to the greater
intensity of the hopping, the system gets a larger contribution of states with
occupation $n \neq 1$. And second, the presence of these states alters the
dynamics, so that for long times the fidelity decreases.

\begin{figure}
  {\centering \resizebox{7cm}{!}{\includegraphics{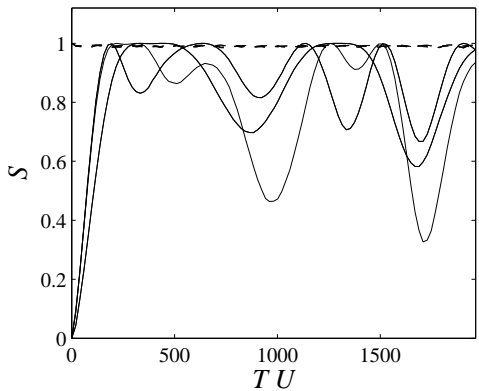}}}
  \caption{\label{fig-Heisenberg}
    Entropy for first, second and third well as a function of time (solid
    lines), for a system with five particles, five wells and open boundary
    conditions, with $J=0.04\, U$ and no external fields. We also plot the
    fidelity (dashed line) with respect to the Heisenberg Hamiltonian
    (\ref{Heisenberg}).}
\end{figure}

Another method to create entanglement in the optical lattice would be not to
use the Ising interaction, but to find an initial state which, under the
influence of the effective Hamiltonian (\ref{Heisenberg}), leads to a larger
amount of entanglement.  The initial state we propose is a set of alternating
spins $\ket{\psi}=\ket{01\ldots0101}$. For two wells, after a time $T=\pi U/
8J^2$, we get the Bell state $\ket{\psi^-}$. For three wells, the states we get
are of the form $\ket{\psi(t)} \propto \left[(1+2e^{i6t})\ket{010} +
  (1-e^{6it})(\ket{100}+\ket{001})\right]/3$. For more than three wells, we
have calculated the states numerically. As we show in Fig.
\ref{fig-Heisenberg}, for open boundary conditions the entropy is not uniformly
distributed and a maximum cannot be reached simultaneously on all wells.
Nevertheless, for a given fidelity, the entanglement of one well with respect
to the others grows faster with the Heisenberg model than with the Ising model.

%------------------------------------------------------------

\section{Quantum simulation and computing with more atoms per well}
In the previous sections we showed how to simulate spin Hamiltonians using
cold atoms in an optical lattice and some external fields. The whole procedure
relied on the assumption that one can load the lattices with precisely one
atom per site. However, this is not the situation of current experiments,
where the number of atoms per cell is small ($1$ to $3$), but otherwise
ignored. Motivated by this limitation, we have developed a scheme for quantum
computation which works independently of how atoms are distributed over the
lattice. As a possible application, this general quantum computer might be
used to simulate spin Hamiltonians.

The method which we summarize in this section has three key ingredients.
First, we have to choose a set of states which are suitable for quantum
computation, called qubits.  Several conditions have to be imposed on the
system to ensure that the different occupation numbers do not influence the
computation. Second, we have to design an effective Hamiltonian which allows
both a predictable interaction between qubits and the possibility to perform
local operations on the qubits by means of external fields. At this point one
realizes that if we do not know how the particles are distributed over the
lattice, then we also lack a lot of information about the Hamiltonians. This
is where the third ingredient comes in: we design a procedure to perform
controlled unitary operations on the lattice, even when most of the parameters
of our system are unknown.

\subsection{Basic tools}

To do quantum computing with $m$ qubits, we need to restrict the possible
configurations of our system to a $2^m$-dimensional Hilbert space, which is
our computation space. Formally, this space is isomorphic to the space of
$\tfrac{1}{2}$-spins developed in Sect. \ref{sec-effective}, but we cannot use
those definitions because they required to have only one atom per lattice
site.  Instead, our implementation of a qubit will rely on the number of atoms
which have been excited from one internal state, $a$, to the other, $b$,
\begin{equation}
  \ket{0} \propto (a^\dagger)^{n}\ket{vac},\;
  \ket{1} \propto b^\dagger (a^\dagger)^{n-1}\ket{vac}.
\end{equation}

In order to ensure the stability of the computation space, we have to impose
certain conditions on our system. First, in the absence of external fields,
the states $\ket{0}$ and $\ket{1}$ should have the same energy. Otherwise,
different configurations of the lattice will acquire different phases which
spoil the computation. To avoid these dephasings, we just need that
$U_{aa}=U_{ab}$. However, for convenience, we will first assume that $U_{aa}$
and $U_{ab}$ are zero, and only at the end study what happens when this is not
true.

The second requirement is that our lattice remains in the computation space at
all times. That is, there can only be at most one atom of type $b$ per site,
and the number of atoms per lattice site must remain constant.  To avoid
exciting more than one atom, there must exist a quantum blockade mechanism
which separates the states $\ket{m}={\bd}^m
{\ad}^{n-m}\ket{vac},\;m=2,3\ldots$, from our computation space and
depopulates them. This blockade is achieved for $U_{bb}\gg U_{ab},U_{aa}$. And
finally, to avoid problems with particles moving from site to site, we will
impose a gradient of energy along the lattice as in Eq.  (\ref{H-el}) with
$\epsilon_k = \epsilon_0 + \delta \times k$. The gradient $\delta$ must be
large compared to $J_b$ and the possible residual values of $J_a$, $U_{aa}$
and $U_{ab}$, but it also has to be smaller than $U_{bb}$ to prevent the
actual motion of atoms.  In other words, $J_{b,a},U_{aa},U_{ab} \ll |U_{bb} -
\delta| \ll \delta, U_{bb}$.  Under the previous conditions, an adiabatic
elimination of the hopping term converts Eq.  (\ref{H-orig}) into an Ising
model
\begin{equation}
  \label{H-cluster}
  H =- \frac{J_b^2}{4(\delta - U_{bb})} \sum_{\langle i,j\rangle}
  (1+\sigma_z^i)(1+\sigma_z^j).
\end{equation}

Apart from an effective interaction between qubits, we also need means to
modify the state of a certain lattice site. To do this we will operate on the
qubits using light or magnetic fields. The interaction between the atoms and
these external fields is governed by the Hamiltonians (\ref{H-mag}) and
(\ref{H-el})
\begin{equation}
  \label{H-light-atom}
  H = \frac{\Delta}{2}\left(\ad_ka_k-\bd_kb_k\right)
  + \frac{\Omega}{2}\left(\ad_kb_ke^{i\phi}+\bd_ka_ke^{-i\phi}\right).
\end{equation}
When we perform an adiabatic elimination of the states outside our computation
space, Eq.  (\ref{H-light-atom}) turns into
\begin{equation}
  \label{H1}
  H_1 = \frac{\Delta}{2}\sigma_z +
  \sqrt{n}\,\frac{\Omega}{2}\left(\sigma_+e^{i\phi}
  +\sigma_-e^{-i\phi}\right).
\end{equation}
This effective Hamiltonian is problematic, because it depends on the
occupation number, which is unknown. It is not be possible for instance to use
a $\pi/2$ pulse to prepare the state $(\ket{0}+\ket{1})/\sqrt{2}$ because,
since do not know $n$, we also cannot predict the time during which the laser
should operate.  Furthermore, one might also argue that, due to the
characteristics of the optical lattice we cannot control the parameters $J_b$,
$U_b$ and $\delta$ with enough precision to do traditional quantum computing
with it.

All previous problems may be formulated in a more general way: we want to
perform accurately certain unitary operations on our set of spins or qubits
using the Hamiltonians $H_1$ (\ref{H1}) and
\begin{equation}
  \label{H2}
  H_2 = \frac{\tilde\Delta}{2}\ket{11}\bra{11},
\end{equation}
even when we do not know some of the constants in them. To be precise, in what
follows we will assume that $\Delta$, $\Omega$, and $\tilde\Delta$ are
unknown, but they can be set to zero and reach a positive value ($\Delta_m$,
$\Omega_m$, and $\tilde\Delta_m$). The only parameter which will be precisely
controlled is the phase of the laser, $\phi$.

There exists a solution to this problem which is based on adiabatic passage.
According to the adiabatic theorem, if we change the parameters of a
Hamiltonian slow enough, and the Hamiltonian has no degenerate eigenstates, we
will be able to perform the unitary transformation $U(T)= \sum_\alpha
e^{i(\phi_\alpha + \psi_\alpha)} |\Phi_\alpha(T)\rangle\langle
\Phi_\alpha(0)|$, where $|\Phi_\alpha(t)\rangle$ are the instantaneous
eigenstates of our system. If the process is designed carefully, and the
geometrical and dynamical phases are properly canceled
($\phi_\alpha=\psi_\alpha=0$), the resulting transformation does not depend on
the precise values of the parameters which governed the evolution ($\Delta$,
$\Omega$, etc), but on the path which the system followed in the space of
possible Hamiltonians. The design of these paths will be discussed in the
following section.

\subsection{Design of the quantum gates}
\label{sec-local}

\begin{figure}[t]
  \resizebox{14cm}{!}{\includegraphics{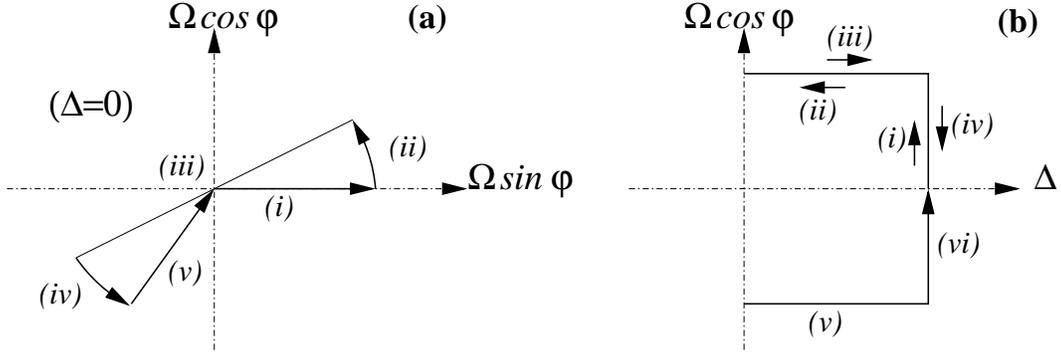}}
  \caption{
    \label{fig-local}
    Schema of how the parameters of Hamiltonian (\ref{H1}) have to be changed
    in order to perform a phase gate (a), and Hadamard gate (b).}
\end{figure}

It is a well known result \cite{Nielsen}, that a quantum computer can be built
upon a small set of unitary operations. By combining these ``gates'' we may
approximate any other transformation as precisely as required.  In this
subsection we will explain how to use the atoms in the optical lattice to
perform a universal set of gates made of a phase gate,
$U=e^{i\theta\sigma_z/2}$, a Hadamard gate and a CNOT gate.

First we care about the phase gate. In (\ref{H1}) we set $\Delta=0$ for all
times and change the remaining parameters $(\Omega,\varphi)$ as depictured in
figure \ref{fig-local}(a):
\begin{eqnarray}
  \label{protocol-phase}
  (0,0) &\stackrel{(i)}{\rightarrow}& (\Omega_m,0)
  \stackrel{(ii)}{\rightarrow} (\Omega_m,\theta/2)
  \stackrel{(iii)}{\Rightarrow} (\Omega_m,\theta/2+\pi) \nonumber \\
  &\stackrel{(iv)}{\rightarrow}& (\Omega_m,\theta+\pi)
  \stackrel{(v)}{\rightarrow} (0,\theta+\pi)
\end{eqnarray}
All steps are followed adiabatically and require a total time $T$, except for
step (iii) whose double arrow indicates a sudden change of parameters.  Note
that $\Omega(0)=\Omega(2T)=0$, $\Omega(t)=\Omega(2T-t)$ and $\varphi(t) = \pi +
\theta - \varphi(2T-t)$, which does not require the knowledge of the function
$f$ but implies a precise control of the phase.  A simple analysis shows that
(i-v) achieve the desired transformation $\ket{0} \to e^{i\theta/2} \ket{0}$,
$\ket{1} \to e^{-i\theta/2} \ket{1}$, with a total cancelation of all
geometrical and dynamical phases.

The Hadamard gate can be realized in a similar way. In the space
$[\Delta,\Omega_x=\Omega\cos(\varphi)]$, the protocol is
\begin{equation}
  \label{protocol-hadamard}
  (0,\Omega_m) \stackrel{(i)}{\rightarrow}
  (\Delta_m,\Omega_m) \stackrel{(ii)}{\rightarrow}
  (\Delta_m,0) \stackrel{(iii)}{\rightarrow}
  (\Delta_m,\Omega_m)
  \stackrel{(iv)}{\rightarrow}
  (0,\Omega_m) \stackrel{(v)}{\Rightarrow}
  (0,-\Omega_m) \stackrel{(vi)}{\rightarrow}
  (\Delta,-\Omega_m) \stackrel{(vii)}{\rightarrow}
  (\Delta,0),
\end{equation}
as shown in figure \ref{fig-local}(b).  In order to avoid the dynamical
phases, we have to make sure that steps (i-v) are run in half the time as
(vi-vii).  More precisely, if $t<T$, we must ensure that
$\Delta(t)=\Delta(T-t)$, $\Omega_x(t)=\Omega_x(T-t)$, $\Delta(T+t) =
\Delta(t/2)$ and $\Omega_x(T+t) = \Omega_x(t/2)$. With this requisite we get
$\frac{1}{\sqrt{2}}(\ket{0} + \ket{1}) \to \ket{0}$,
$\frac{1}{\sqrt{2}}(\ket{0} - \ket{1}) \to -\ket{1}$.  Again, the whole
procedure does not require us to know $\Omega$ or $\Delta$, but rather to
control the evolution of the experimental parameters which determine them.

For the last universal gate we will take a pair of interacting qubits, and
shine a laser on one of them. The total effective Hamiltonian may be written as
\begin{equation}
  H = \frac{\tilde\Delta}{2}\ket{11}\bra{11}
  + \one \otimes \frac{\Omega}{2}(\sigma_+e^{i\phi}+\sigma_-e^{-i\phi}).
\end{equation}
By changing the parameters $[\tilde\Delta,\Omega_x=\Omega\cos(\varphi)]$ in the
following way,
\begin{equation}
  \label{protocol-nonlocal}
  (\tilde\Delta_m,0) \stackrel{(i)}{\rightarrow}
  (\tilde\Delta_m,\Omega_m)
  \stackrel{(ii)}{\rightarrow} (0,\Omega_m)
  \stackrel{(iii)}{\Rightarrow} (0,-\Omega_m)
  \stackrel{(iv)}{\rightarrow} (\tilde\Delta_m,-\Omega_m)
  \stackrel{(v)}{\rightarrow} (\tilde\Delta_m,0).
\end{equation}
we obtain the transformation
\begin{equation}
  U_1 = \ket{0}\bra{0}\otimes \one + e^{i\xi}\ket{1}\bra{1}\otimes i\sigma_y,
\end{equation}
where $\xi = \int_0^T \delta(t) dt$ is an unknown dynamical phase. To get rid
of this phase we have to apply two more unitaries. First we need to perform a
NOT on the first qubit $U_2 = (\ket{0}\bra{1} + \ket{1}\bra{0}) \otimes \one$.
Finally, if $\tilde\Delta^{(1)}(t)$ denotes the evolution of $\tilde\Delta$ in
equation (\ref{protocol-nonlocal}), the system must follow a path such that
$\tilde\Delta^{(3)}(t) = \tilde\Delta^{(1)}(t)$, $\Omega^{(3)}(t) = 0$.  If
the timing is correct, we achieve $U_3 = (\ket{0}\bra{0} +
e^{i\xi}\ket{1}\bra{1})\otimes \one$.  Everything combined gives us the CNOT
up to a global phase $U_{cnot} = \ket{0}\bra{0}\otimes \one
+\ket{1}\bra{1}\otimes i\sigma_y = e^{-i\xi}U_2 U_3 U_2 U_1$, which does not
affect the computation.

\subsection{Errors}

In our design for a quantum computer, there are many implicit approximations,
which in practice become sources of error. Basically, we have neglected
processes which lead to (i) excitation of more than one atom into state
$\ket{b}$, (ii) change of occupation numbers due to hopping of atoms, and
(iii) hopping and interaction of atoms in state $\ket{a}$. The first two
phenomena are suppressed if $(\Omega/U_{bb})^2\ll 1$ and
$(J^{(a)}_k/U_{bb})^2\ll 1$. We may analyze the remaining errors
perturbatively, and study how they modify the effective Hamiltonians
(\ref{H1}) and (\ref{H2}).  First, the the virtual excitation of two atoms
increments the parameter $\Delta$ by an unknown amount, $\Delta_{eff} \sim
\Delta+ 2\Omega^2n_k / (\Delta+U_{ab}-U_{bb})$. But if $U_{ab}\ll U_{bb}$ and
$\Omega^2n_k T/U_{bb} \ll 1$, this shift may be neglected.  And second, in the
two-qubit Hamiltonian (\ref{H2}) there appear additional contributions due to
virtual hopping of all types of atoms, which are of order
$\max(J_b,J_a)^2/\delta^2 \sim J^2/U_{bb}$.  For $J^2T/U_{bb} \ll 1$ these
energy shifts may also be neglected.

\begin{figure}[t]
  \centering
  \resizebox{\linewidth}{!}{\includegraphics{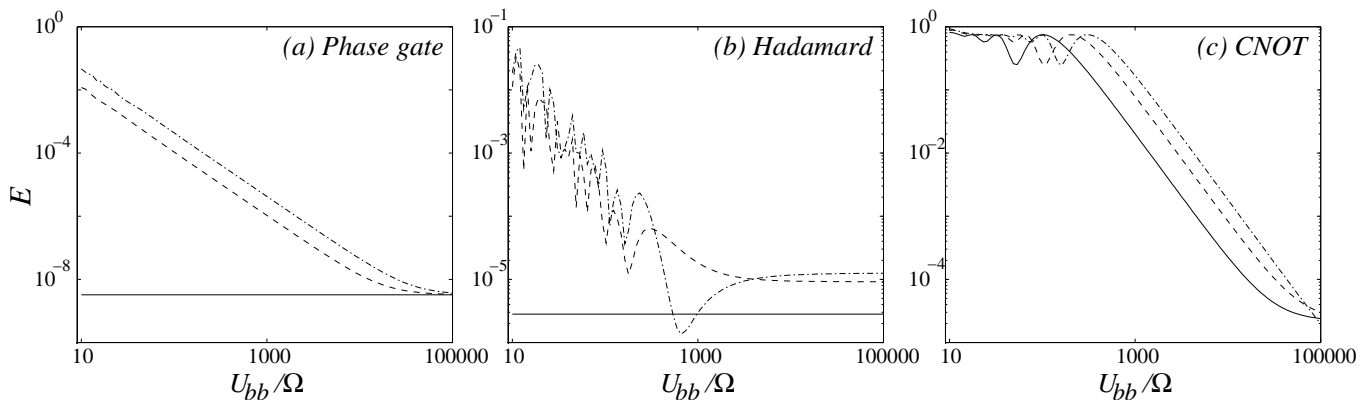}}
  \caption{\label{fig-error2}
    Log-log plot of the gate error, $E=1-{\cal F}$, for the (a) phase, (b)
    Hadamard and (c) CNOT gates. The parameters for the simulations are
    $U_{aa}=U_{ab}=J_m$, $J_a=J_b$,$g=U_{bb}+U_{ab}/2$, and
    $T=100/\Omega_m$. For the local gates we choose $\Delta_m=6\Omega_m=1$,
    and for the nonlocal gate $\Omega_m=J_m^2/6$. For each simulation we
    choose different population imbalance ($|n-m|=0,1,2$ for solid, dashed
    and dotted lines), and change the interaction constant $U_{bb}$.}
\end{figure}

To check the validity of our approximations, we have simulated the evolution
of two atomic ensembles with an effective Hamiltonian which results of
applying second order perturbation theory to Eq. (\ref{H-orig}), and which
takes into account all important processes. As a figure of merit we have
chosen the gate fidelity \cite{Nielsen}
\begin{equation}
{\cal F} = 2^{-n}|\mathrm{Tr}\{U_{ideal}^\dagger U_{real}\}|^2
\end{equation}
where $n$ is the number of qubits involved in the gate, $U_{ideal}$ is the gate
that we wish to produce and $U_{real}$ is the actual operation performed.  The
results are plotted in Fig. \ref{fig-error2}. In these pictures we show the
error of the gates for simulations in which all parameters are fixed, except
for $U_{bb}$ and the occupation numbers of the wells. The first conclusion is
that the stronger the interaction between atoms in state $\ket{b}$, the smaller
the energy shifts.  This was already evident from our analytical estimates,
because all errors are proportional to $1/U_{bb}$.  Typically, a ratio $U_{bb}
= 10^4 U_{ab}$ is required to make ${\cal F} = 1-10^{-4}$, but reasonable
fidelities may be achieved for more realistic values.  Finally, the larger the
number of atoms per well, the poorer the fidelity of the local gates [Figure
\ref{fig-error2}(a-b)].

%------------------------------------------------------------

\section{Conclusions}
We have shown that a Mott insulator of bosonic atoms with filling factor one
and two accessible internal states behaves as a lattice of localized spins.
This result generalizes that of \cite{Duan}, by considering not only atoms in
an optical lattice, but also the influence of external fields, such as lasers,
magnetic fields, Rabi oscillations, etc. As we showed, tuning the parameters of
the potential which confines the atoms and using additional elements, such as a
magnetic field or lasers, the spins may be forced to simulate a great variety
of Hamiltonians, including the Heisenberg and Ising models, with either
attractive ($U_{ab}>0$) or repulsive ($U_{ab}<0$) interactions. When, as in
current experiments, the number of atoms per lattice site is unknown ---for
instance there is some superfluid component, the filling factor is not an
integer or there is some external potential---, we may still draw the atoms
into a regime in which they behave as localized spins, but we will not know the
intensity of the interactions of the system with external fields.
Nevertheless, even ignoring the parameters which govern their dynamics, we have
shown that it is possible to perform a universal set of gates and do quantum
computation.

\acknowledgments

We thank D. Liebfried and P. Zoller for discussions and the EU project EQUIP
(contract IST-1999-11053).

\end{document}